\documentclass[prl,a4paper]{revtex4}
\usepackage{epsfig}
\usepackage{amsfonts}
\usepackage{amsmath}
\usepackage{graphicx}

\newcommand{\eq}{\begin{equation}}
\newcommand{\eqx}{\end{equation}}
\newcommand{\eqn}{\begin{eqnarray}}
\newcommand{\eqnx}{\end{eqnarray}}
\newcommand{\f}[2]{\frac{#1}{#2}}

\newcommand{\al}{\alpha}
\newcommand{\bt}{\beta}

\newcommand{\eps}{\varepsilon}
\newcommand{\nn}{{\cal N}}
\newcommand{\arctanh}{\mbox{\rm arctanh}\,}

\newcommand{\cor}[1]{\left\langle{#1}\right\rangle}

\begin{document}

\title{Viscous plasma evolution from gravity using AdS/CFT}  

\author{Romuald A. Janik}\email{ufrjanik@if.uj.edu.pl}
\affiliation{Institute of Physics, Jagellonian University,
Reymonta 4, 30-059 Krakow, Poland.}

\begin{abstract}
We analyze the AdS/CFT dual geometry of an expanding boost-invariant plasma.
We show that the requirement of nonsingularity of the dual geometry
for leading and subasymptotic times predicts, without any further
assumptions about gauge theory dynamics, hydrodynamic expansion
of the plasma with viscosity coefficient exactly matching the one obtained
earlier in the static case by Policastro, Son and Starinets.   
\end{abstract}

\maketitle

One of the more challenging problems in theoretical physics is the
understanding from first principles of the behaviour of quark-gluon
plasma, the phase of matter consisting of deconfined quarks and
gluons. There are strong indications that the plasma observed at
RHIC is indeed strongly coupled (see e.g. \cite{review}) and is well
described by models based on hydrodynamics \cite{hydro}. This suggests
the need for nonperturbative techniques to study the dynamics of the
system. This need is especially acute if one would like to study
non-equilibrium phenomena, thermalization etc.

A very powerful technique for studying nonperturbative properties of
gauge theory has emerged from string theory in recent years
\cite{adscft}. In its original form the AdS/CFT
correspondence states the
equivalence of $\nn=4$ Super Yang-Mills theory and string theory in a
curved 10-dimensional $AdS_5 \times S^5$ background. What is crucial is
that the string theory is easiest to handle in the regime of strong
gauge theory coupling.

Although the correspondence does not have
a direct counterpart which works for QCD it has been argued that
for studying features of the plasma not too far above the deconfinement phase
transition it may be a good approximation as the plasma is strongly
coupled and deconfined.

From a more general perspective, it is also interesting to study, for
their own sake, 
dynamical time-dependent processes in the $\nn=4$ supersymmetric
theory within the context of the AdS/CFT correspondence in order to
have an example where such phenomenae can be calculated exactly, as
well as to develop in this context new methods to address these
dynamical issues. The methods used in the present paper translate
questions on the behaviour of the plasma into certain questions within
general relativity and we hope that this may inspire further research
in both domains.   

Properties of the $\nn=4$ gauge theory at fixed finite temperature
have been studied 
in some detail \cite{son,other}, in particular shear viscosity has
been calculated from perturbations around a static black hole
background. 
On a more qualitative level, thermalization has been suggested to
correspond to black hole formation in the dual description
\cite{nastase}, while cooling was advocated to correspond to black
hole motion in the 5th direction \cite{zahed}.

In \cite{US1} a quantitative framework for studying such
time-dependent phenomena has been proposed. The criterion of a {\em
nonsingular} dual geometry was shown to pick out uniquely, in a boost-invariant
setting, asymptotic perfect fluid hydrodynamical evolution for large
proper-times. The resulting asymptotic geometry was shown to be
analogous to a moving black hole. Further work within this framework
includes \cite{US2,nakamura}.

The aim of this paper is to show that the criterion of nonsingularity
predicts, when applied also to subasymptotic times, {\em viscous}
hydrodynamic evolution with a specific viscosity coefficient. As a
byproduct we obtain a nontrivial consistency check of the AdS/CFT
correspondence for the value of the shear viscosity.

{\bf Boost-invariant viscous hydrodynamics. }
Let us consider the spacetime evolution of the
energy-momentum tensor of an expanding fireball of plasma. The
energy-momentum tensor is constrained by energy-momentum conservation
\eq
\partial_\mu T^{\mu\nu}=0
\eqx
and, for the case of $\nn=4$ SYM theory that we consider here,
tracelessness $T^\mu_\mu=0$. We will further restrict ourselves to
boost-invariant evolution, first considered by Bjorken \cite{Bjorken}
as a model of the mid-rapidity region in heavy-ion collisions. This
assumption is also commonly used in hydrodynamic simulations for
RHIC \cite{hydro}. We will further assume no dependence on transverse
coordinates. It is natural to use the proper-time/spacetime rapidity
coordinates for Minkowski space:
\eq
ds^2=-d\tau^2+\tau^2 dy^2+dx_\perp^2
\eqx 
Then energy-conservation and tracelessnes determine $T^{\mu\nu}$ in
terms of a single function -- the energy density $\eps(\tau)$ (for
explicit expressions~\footnote{In that paper $f(\tau) \equiv
  \eps(\tau)$.} see \cite{US1}). The dynamics of gauge theory should
then determine $\eps(\tau)$. In \cite{US1} we proposed to use AdS/CFT
to determine the proper-time dependence of energy density by first
constructing the dual geometry to a given $\eps(\tau)$ and then
requiring its nonsingularity to fix the physical $\eps(\tau)$.
 
Let us review what would be the physical expectations for {\em large}
proper-times if the gauge
theory dynamics were described by viscous hydrodynamics.
If there would be no viscosity, then we would have
\eq
\label{e.bjorken}
\eps(\tau)=\f{1}{\tau^{\f{4}{3}}}
\eqx
If in addition we would have viscosity~\footnote{The power of $\tau$
  represents $\eta \propto T^3$.} 
\eq
\eta=\f{\eta_0}{\tau}
\eqx
then the energy density would behave like (see \cite{nakamura})
\eq
\label{e.gauge}
\eps(\tau)=\f{1}{\tau^{\f{4}{3}}}-\f{2\eta_0}{\tau^2}+\ldots
\eqx
with viscosity effects generating subleading deviations from the ideal
fluid case (\ref{e.bjorken}).

{\bf Construction of a dual AdS/CFT geometry. }
The procedure of constructing the dual geometry to a gauge theory
configuration with given expectation value of the energy-momentum
tensor was introduced in \cite{Skenderis}. One adopts the
Fefferman-Graham coordinates \cite{fg} for the 5-dimensional metric
\eq
ds^2=\f{\tilde{g}_{\mu\nu} dx^\mu dx^\nu + dz^2}{z^2}
\eqx
where the $z$ coordinate is the `fifth' coordinate while $\mu$ is a 4D
index. One solves Einstein equations with negative cosmological
constant:
\eq
\label{e.einst}
E_{\al\bt} \equiv R_{\al\bt}-\f{1}{2} g_{\al\bt} R-6g_{\al\bt}=0
\eqx
with a boundary condition for $\tilde{g}_{\mu\nu}$ around $z=0$:
\eq
\label{e.bc}
\tilde{g}_{\mu\nu}=\eta_{\mu\nu}+z^4 \tilde{g}^{(4)}_{\mu\nu}+\ldots
\eqx
The fourth order term is related to the expectation value of the
energy-momentum tensor through
\eq
\label{e.tmunu}
\cor{T_{\mu\nu}}=\f{N_c^2}{2\pi^2} \tilde{g}^{(4)}_{\mu\nu}
\eqx
The procedure is therefore to solve the 5-dimensional Einstein's
equations (with negative cosmological constant $\Lambda=-6$) with the
boundary condition (\ref{e.bc}) for a given 
spacetime profile of the gauge-theoretical $\cor{T_{\mu\nu}}$.
In the following since we will be dealing directly with the metric and
hence with $\tilde{g}^{(4)}_{\mu\nu}$ so we will supress the factor
$N_c^2/(2\pi^2)$ throughout the computation reinstating it only in the
final discussion. 

{\bf Nonsingularity and viscosity. }
The metric consistent with the symmetries of the boost invariant
expansion has the form
\eq
ds^2=\f{-e^{a(\tau,z)}d\tau^2+\tau^2 e^{b(\tau,z)} dy^2+e^{c(\tau,z)}
  dx_\perp^2+dz^2}{z^2} 
\eqx
In \cite{US1} the dual geometry was determined for asymptotic times
with the energy-density behaving like
\eq
\eps(\tau)=\f{1}{\tau^s}
\eqx
for large proper-times.
The resulting metric coefficients were given for large proper-times
as functions of the scaling variable
\eq
v=\f{z}{\tau^{\f{s}{4}}}
\eqx
It was found in \cite{US1} that the resulting geometry was nonsingular
only when $s=4/3$, thus corresponding to perfect fluid
hydrodynamics. Of course subleading corrections like (\ref{e.gauge})
are possible. The explicit leading coefficients for this case were found to be
\eqn
a(v)&=&\log \f{(1-v^4/3)^2}{1+v^4/3} \nonumber \\
\label{e.leading}
b(v)&=&c(v) = \log (1+v^4/3)
\eqnx

The resulting geometry looks like a black hole whose horizon (in the
Fefferman-Graham coordinates) moves in the fifth dimension as
$z_0=3^{\f{1}{4}} \tau^{\f{1}{3}}$. This leads to the temperature 
\eq
\label{e.temp}
T=\f{\sqrt{2}}{\pi} \f{1}{z_0}=\f{\sqrt{2}}{\pi 3^{\f{1}{4}}} \tau^{-\f{1}{3}}
\eqx
The coefficient $\sqrt{2}$ comes from the special form of the black
hole metric in Fefferman-Graham coordinates. See \cite{US1} for a
discussion. 
 
In an interesting recent paper Nakamura and Sin \cite{nakamura}
determined the leading, subasymptotic in proper-time, corrections to
the metric coefficients like 
\eq
a(\tau,z)=a(v)+a_1(v) \f{1}{\tau^{\f{2}{3}}}+\ldots
\eqx
which represent $\eps(\tau)$ of the form (\ref{e.gauge}). The
coefficients which can be extracted from the results of
\cite{nakamura} in a form convenient for proceeding to higher orders
are
\eqn
a_1(v) &=& 2\eta_0 \f{(9+v^4)v^4}{9-v^8} \nonumber \\
b_1(v) &=& -2 \eta_0 \f{v^4}{3+v^4} +2\eta_0 \log \f{3-v^4}{3+v^4}
\nonumber \\
\label{e.first}
c_1(v) &=& -2 \eta_0 \f{v^4}{3+v^4} -\eta_0 \log \f{3-v^4}{3+v^4}
\eqnx

However to this order in $\tau$ (${\cal O}(\tau^{-\f{2}{3}})$) the Riemann
tensor squared
\eq
\label{e.rsquare}
{\mathfrak R}^2 \equiv R^{\mu\nu\al\bt}R_{\mu\nu\al\bt}  
\eqx
was found to be finite for {\em any} $\eta_0$, the
first divergence appearing only at order $\tau^{-\f{4}{3}}$. This
behaviour suggested
that in order to determine the coefficient of viscosity $\eta_0$ one had to go
one order higher in $\tau^{-\f{2}{3}}$. Namely one has to find the
metric coefficients to second order like
\eq
a(\tau,z)=a(v)+a_1(v) \f{1}{\tau^{\f{2}{3}}}+a_2(v)
\f{1}{\tau^{\f{4}{3}}}+ \ldots
\eqx 
In order to setup a systematic expansion procedure
one expands the left hand side of the Einstein equations
$E_A \equiv (\tau^{\f{2}{3}} E_{\tau\tau},\tau^{\f{4}{3}} E_{\tau z},
\tau^{\f{2}{3}} E_{zz}, \tau^{-\f{4}{3}} E_{yy}, \tau^{\f{2}{3}}
E_{xx})$ in powers of $\tau^{-\f{2}{3}}$:
\eq
E_A=E_A^{(0)}(v) + E_A^{(1)}(v) \f{1}{\tau^{\f{2}{3}}} +E_A^{(2)}(v)
\f{1}{\tau^{\f{4}{3}}} +\ldots 
\eqx
where $A=1\ldots 5$ numbers the five nontrivial components of the
Einstein equations mentioned above.
The prefactors are chosen in
such a way as to have a uniform expansion. The leading order results
(\ref{e.leading}) satisfy $E_A^{(0)}=0$ while the first subleading
corrections (\ref{e.first}) satisfy $E_A^{(1)}=0$.
Therefore in order to find
$a_2(v),b_2(v),c_2(v)$ we have to solve the equations $E_A^{(2)}=0$
with the boundary condition that these coefficient functions should
vanish at $v=0$. The resulting expressions depend on the viscosity
coefficient $\eta_0$ and another coefficient which we denote by
$C$. The explicit expressions are
\eqn
\!\!\!\!a_2(v)\!\! &=&\!\! \int_0^v \left(
\f{4w^5(27+9w^4+2w^8)}{3(9-w^8)^2} -24 \eta_0^2 
\f{w^3 (405+171w^4+189w^8+5w^{12}-2w^{16})}{(9-w^8)^3} -C \f{w^3
  (9+2w^4+w^8)}{(9-w^8)^2} \right)  dw \nonumber \\
\!\!\!\!b_2(v)\!\! &=&\!\! -2c_2(v)+\f{v^2}{6+2v^4}-\f{\arctanh
  \f{v^2}{\sqrt{3}}}{2\sqrt{3}} +\eta_0^2 \f{v^4 (39+7v^4)}{(3+v^4)^2}
+\f{3}{2} \eta_0^2 \log \f{3-v^4}{3+v^4} +C \f{v^4}{12(3+v^4)}
\nonumber \\
\!\!\!\!c_2(v)\!\! &=&\!\! \int_0^v \left( \f{4w^9(9+w^4)}{3(9-w^8)^2} +\f{8w^3
  \arctanh \f{w^2}{\sqrt{3}}}{\sqrt{3} (9-w^8)} +\eta_0^2 \f{24 w^3
  (w^4-15)(3+5w^4)}{(3-w^4)^2 (3+w^4)^3} -C
\f{w^3(1+w^4)}{(3-w^4)(3+w^4)^2} \right) dw
\eqnx
One can express these integrals in terms of elementary functions and
dilogarithms but the expressions are rather lengthy and will not be presented
here.

It remains to determine when the background geometry given by the
coefficients up to second order is nonsingular. To this end we
calculate the curvature invariant ${\mathfrak R}^2$ defined by
(\ref{e.rsquare}) 
and expand it in the scaling limit up to the order
$\tau^{-\f{4}{3}}$. The resulting expression at this order has the
form
\eq
{\mathfrak R}^2=\mbox{\rm nonsingular terms} +\f{1}{\tau^{\f{4}{3}}}
\f{\mbox{\rm polynomial in $v$, $\eta_0$ and $C$}}{(3-v^4)^4 (3+v^4)^6}
\eqx
We see that there is a potential singularity at $v=3^{\f{1}{4}}$. It
turns out that the singularity is cancelled exactly when
\eq
\label{e.eta0}
\eta_0^2= \f{\sqrt{3}}{18}
\eqx
There is no restriction on $C$ at this order. We expect one would
have to perform the analysis to the next order to fix $C$ \cite{WIP}.

The above calculation shows that nonsingular dual geometry is not possible
for an exact perfect fluid, but that viscosity effects in the proper-time
evolution are present with a uniquely fixed value of the viscosity
coefficient given by (\ref{e.eta0}). It is interesting to compare this
value with the shear viscosity obtained by Policastro, Son and Starinets
\cite{son} who derived it by studying the reponse of a static plasma at fixed
temperature to small perturbations.

To this end let us take the value of (shear) viscosity obtained in \cite{son}
at a fixed temperature $T$:
\eq
\eta=\f{1}{4\pi} s=\f{\pi}{8} N_c^2 T^3
\eqx
If we insert the proper-time dependence of the temperature
(\ref{e.temp}) for the evolving plasma into the
above expression we obtain
\eq
\eta=\f{N_c^2}{2\pi^2} \cdot \f{1}{2^{\f{1}{2}} 3^{\f{3}{4}}} \cdot
\f{1}{\tau} 
\eqx
where we have factored out the coefficient $N_c^2/(2\pi^2)$ appearing in
(\ref{e.tmunu}).
The resulting estimate for the viscosity coefficient $\eta_0$ is
therefore
\eq
\eta_0= \f{1}{2^{\f{1}{2}} 3^{\f{3}{4}}} \equiv \left(
\f{\sqrt{3}}{18}  \right)^{\f{1}{2}}
\eqx
which is exactly the value (\ref{e.eta0}) for which the dual
background geometry of the evolving plasma is nonsingular.

{\bf Discussion. }
In this paper we have studied the proper-time evolution of a
boost-invariant plasma using the AdS/CFT correspondence. We have
shown that the requirement that the dual geometry is nonsingular
predicts the proper-time evolution of the energy density to be
equal~\footnote{Up to the considered order in $1/\tau$.} to the one
found from viscous 
hydrodynamics with the viscosity being exactly the one following from
$\eta/s=1/(4\pi)$ in the static case. The computation involves the
nonlinear regime of gravity on the AdS/CFT side and it is encouraging
for other possible applications that the method can capture such fine
details of the gauge theory dynamics.

It would be very interesting to study in more detail the features of
this geometry and its thermodynamics \cite{WIP}, as well as to apply
these techniques to other dynamical non-equilibrium processes.

\bigskip

\noindent{}{\bf Acknowledgments.} This work has been completed during
the program `From RHIC to LHC: Achievements and Opportunities' at the
Institute of Nuclear Theory, Seattle. I would like to thank the INT
for hospitality. This work has been supported in part by Polish
Ministry of Science and Information Society Technologies grants
1P03B02427 (2004-2007), 1P03B04029 (2005-2008) and RTN network ENRAGE
MRTN-CT-2004-005616.

\end{document}